\documentclass{jetpl}
\twocolumn

\usepackage{amsmath,amssymb,epsfig,color,pstricks,bm,graphics,alltt}

\begin{document}

\lat

\title{Electronic Structure of Prototype AFe$_2$As$_2$ and ReOFeAs 
High-Temperature Superconductors: a Comparison}

\rtitle{Electronic Structure of AFe$_2$As$_2$ and ReOFeAs}

\sodtitle{Electronic Structure of Prototype AFe$_2$As$_2$ and ReOFeAs  
High-Temperature Superconductors:  a Comparison}

\author{I.\ A.\ Nekrasov$^+$, Z.\ V.\ Pchelkina$^*$, M.\ V.\ Sadovskii$^+$}

\rauthor{I.\ A.\ Nekrasov, Z.\ V.\ Pchelkina, M.\ V.\ Sadovskii}

\sodauthor{Nekrasov, Pchelkina, Sadovskii }

\sodauthor{Nekrasov, Pchelkina, Sadovskii }

\address{$^+$Institute for Electrophysics, Russian Academy of Sciences, 
Ural Division, 620016 Ekaterinburg, Russia}

\address{$^*$Institute for Metal Physics, Russian Academy of
Sciences, Ural Division, 620041 Ekaterinburg GSP-170, Russia}

\dates{Today}{*}

\abstract{
We have performed {\it ab initio} LDA calculations of electronic 
structure of newly discovered prototype high-temperature superconductors  
AFe$_2$As$_2$ (A=Ba,Sr) and compared it with previously calculated
electronic spectra of ReOFeAs (Re=La,Ce,Pr,Nd,Sm). 
In all cases we obtain almost identical densities of states in rather wide energy interval 
(up to 1 eV) around the Fermi level. Energy dispersions are also very similar
and almost two-dimensional in this energy interval, leading to the same basic 
(minimal) model of electronic spectra, determined mainly by Fe 
$d$-orbitals of FeAs layers.
The other  constituents, such as A ions or rare earths Re (or oxygen states)
are more or less irrelevant for superconductivity.
LDA Fermi surfaces for AFe$_2$As$_2$ are also very similar to that of
ReOFeAs. 
This makes the more simple AFe$_2$As$_2$ a generic system to study 
high-temperature superconductivity in FeAs - layered compounds.}


\maketitle

The recent discovery of the new superconductor LaO$_{1-x}$F$_x$FeAs with the
transition temperature $T_c$ up to 26K \cite{kamihara_08,chen,zhu,mand} 
and even more high values of $T_c=$41-55K in CeO$_{1-x}$F$_x$FeAs \cite{chen_3790}, 
SmO$_{1-x}$F$_x$FeAs \cite{chen_3603},  NdO$_{1-x}$F$_x$FeAs and 
PrO$_{1-x}$F$_x$FeAs \cite{ren_4234,ren_4283} was recently followed by the
discovery of high-temperature superconductivity with $T_c$ up to 38K in K doped
ternary iron arsenides BaFe$_2$As$_2$ \cite{rott} and SrFe$_2$As$_2$ 
\cite{ChenLi}, with further synthesis of superconducting AFe$_2$As$_2$ 
(A=K, Cs, K/Sr, Cs/Sr) \cite{Chu}. Relatively large single crystals of 
superconducting Ba$_{1-x}$K$_x$Fe$_2$As$_2$ were also grown \cite{Bud},
providing a major breakthrough in the studies of anisotropic electronic
properties of FeAs - layered superconductors.  

The LDA electronic structure of LaOFeAs were 
calculated in a number of papers (see e.g. \cite{singh}, \cite{dolg},
\cite{mazin}) producing results qualitatively similar to that first obtained
for LaOFeP \cite{lebegue}. We have performed LDA calculations for the whole
series of ReOFeAs (R=La,Ce,Pr,Nd,Sm) \cite{Nekr}, demonstrating a very weak (or
absent) dependence of electronic spectrum on the type of the rare-earth ion Re
in rather wide energy interval ($\sim 2$eV) around the Fermi level.

First LDA results for the density of states (DOS) of BaFe$_2$As$_2$ were
recently presented in Refs. \cite{Shein, Krell}.
Here we present the results of our {\it ab initio} calculations of 
electronic structure of the newly discovered prototype high-temperature 
superconductors AFe$_2$As$_2$ (A=Ba, Sr) with the aim to compare it with the
previously discussed ReOFeAs series. We present LDA DOS, energy dispersions and 
Fermi surfaces of these compounds and briefly discuss possible conclusions
with respect to the minimal model of electronic spectrum and superconductivity.
As all results are quite similar for both A=Ba and A=Sr, as well as for the 
whole Re series, below we present data mainly for A=Ba and Re=La.


Both BaFe$_2$As$_2$ and LaOFeAs crystallize in tetragonal structure with the 
space group $I$4/$mmm$ and $P$4/$nmm$, correspondingly. Both compounds are formed 
of (FeAs)$^{-}$ layers alternating with Ba$_{0.5}^{2+}$ or (LaO)$^{+}$. Fe$^{2+}$ 
ions are surrounded by four As ions forming a tetrahedron. The crystal structures 
of BaFe$_2$As$_2$ and LaOFeAs are shown in Fig. 1.
The quasi two-dimensional character of both compounds makes them similar to the 
well studied class of superconducting copper oxides. At 140 K BaFe$_2$As$_2$ 
undergoes structural phase transition from tetragonal ($I$4/$mmm$) to 
orthorhombic ($Fmmm$) space group~\cite{rotter_4021}. The same transition takes 
place for LaOFeAs system at 150 K: 
$P$4/$nmm$ (tetragonal)$\rightarrow$ $Cmma$ (orthorhombic)~\cite{Nomura_3569}. 
The crystallographic data for tetragonal phase of two compounds is collected 
in Table 1.
It can be seen that for BaFe$_2$As$_2$ compound the 
Fe-As distance is smaller than for LaOFeAs. So one would expect more 
considerable Fe-$d$-As-$p$ hybridization for BaFe$_2$As$_2$ system in comparison 
with LaOFeAs and as a result wider Fe-$d$ bandwidth.
The distance between nearest Fe atoms within FeAs layers is also significantly 
smaller in BaFe$_2$As$_2$ as compared with LaOFeAs system. After the phase 
transition of BaFe$_2$As$_2$ system to the orthorhombic structure the four equal 
Fe-Fe distances break into two bond pairs of 2.808 \AA~and 2.877 \AA~length. 
Moreover the two As-Fe-As angles are quite different in the case of LaOFeAs 
system (113.6$^\circ$ and 107.5$^\circ$) and have very close values 
($\sim 109^\circ$) for BaFe$_2$As$_2$. Such differences in the nearest 
surrounding of Fe ions should evoke the distinctions in the electronic structure 
of these two compounds.

\begin{figure}[!h]
\includegraphics[clip=true,width=0.2\textwidth]{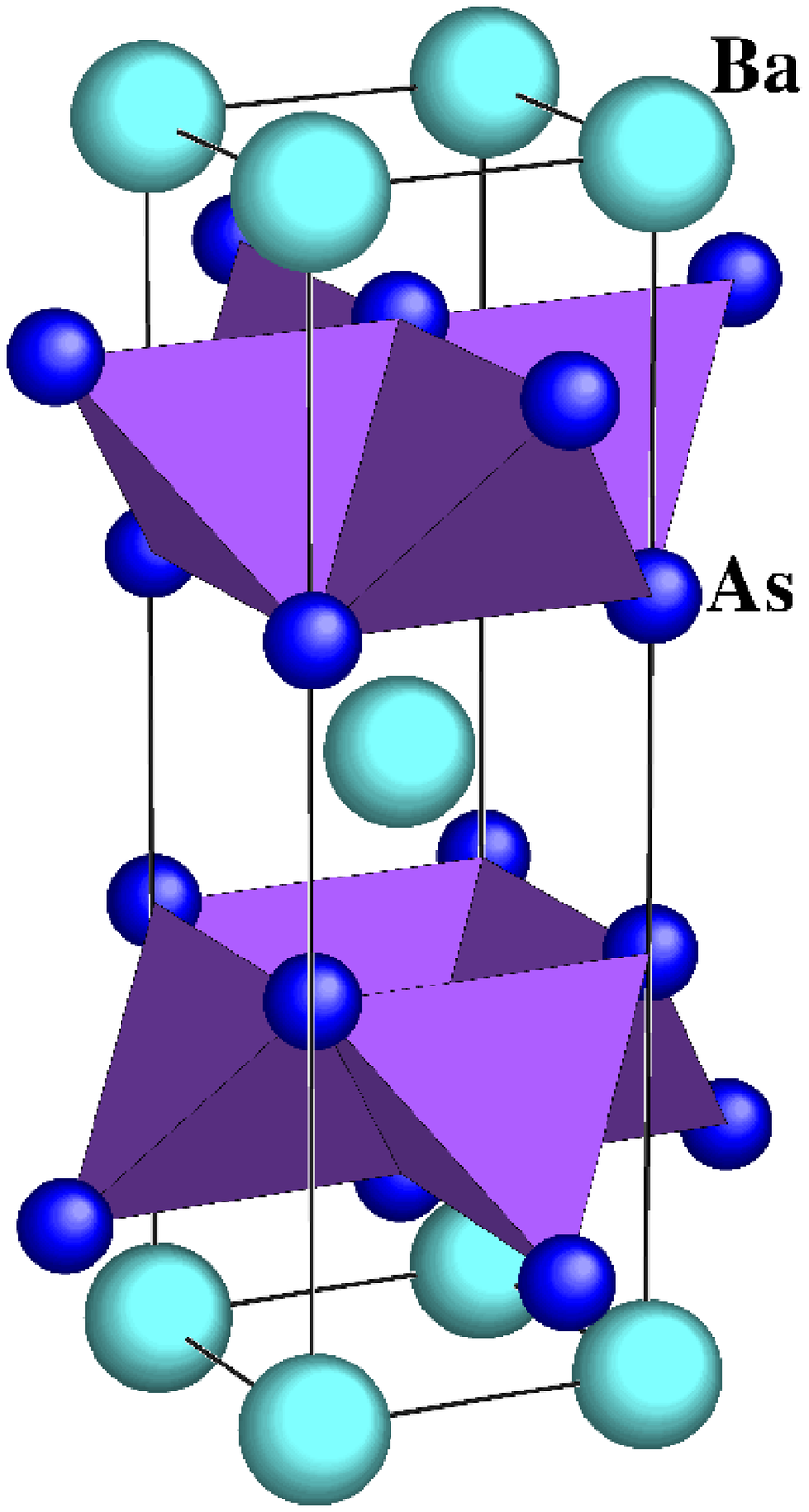}
\includegraphics[clip=true,width=0.2\textwidth]{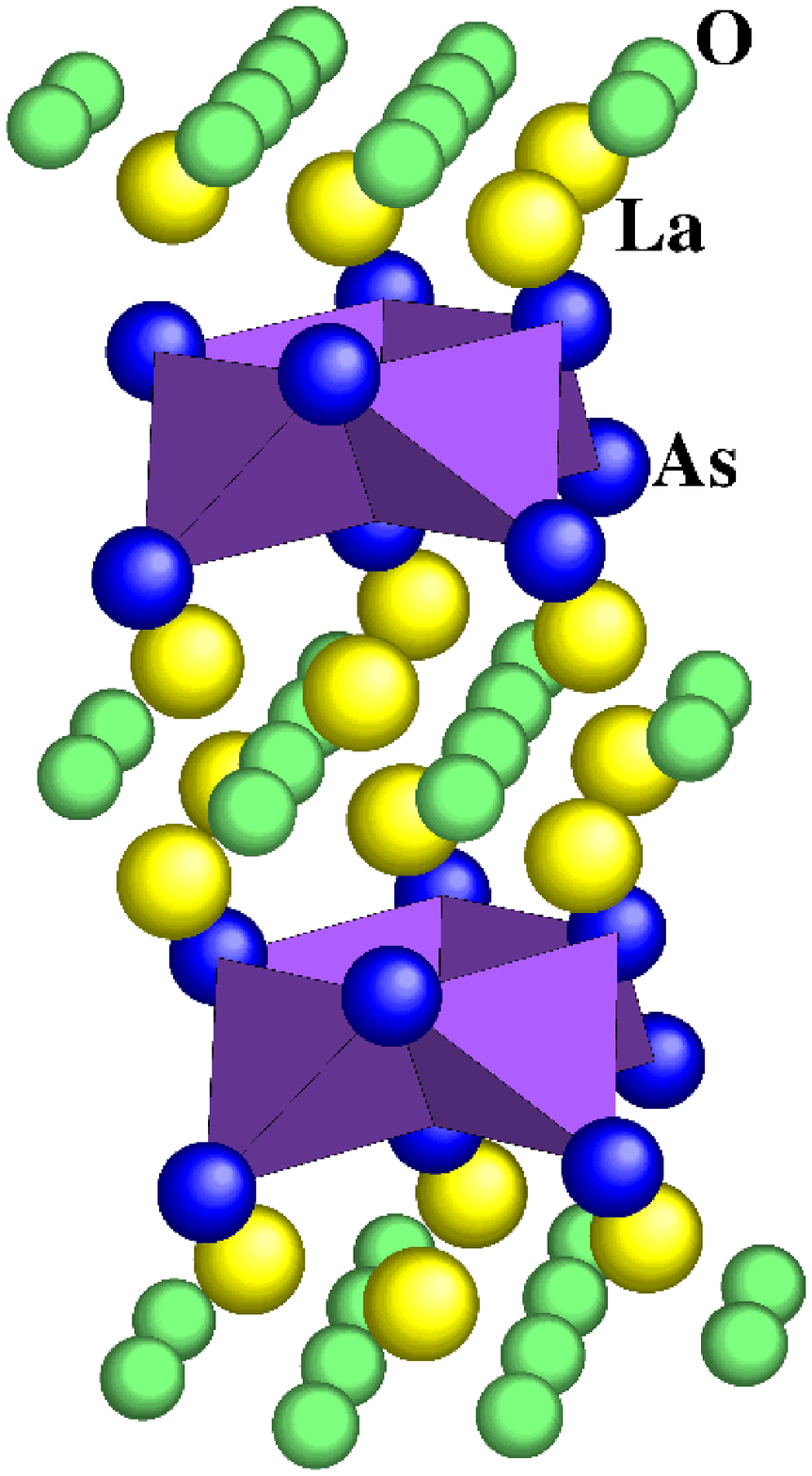}
\label{ba_la_struc}
\caption{Fig. 1. Crystal structure of BaFe$_2$As$_2$ (left) and LaOFeAs (right). 
FeAs tetrahedra (violet) form two-dimensional layers sandwiched by Ba ion (cyan) or LaO layers (yellow and green).}
\end{figure}
\begin{table}[htb]
\caption{
Table 1.
Crystal structure data for BaFe$_2$As$_2$ and LaOFeAs compounds. 
Atomic positions for BaFe$_2$As$_2$ are Ba (0, 0, 0), Fe (0.5, 0, 0.25), As (0, 0, $z$) and for LaOFeAs are La(0.25, 0.25, $z$), Fe (0.75, 0.25, 0.5), As (0.25, 0.25, $z$), O (0.75, 0.25, 0).}
\label{tab1}
\begin{tabular}{|l|c|c|}
\hline
Parameter &BaFe$_2$As$_2$&LaOFeAs \\   
\hline
group     &I4/mmm	   & P4/nmm  \\  
$a$, \AA  & 3.9090(1)    &4.03533(4) \\
$c$, \AA  &13.2122(4)    &8.74090(9) \\
$z_{La}$  &  -           &0.14154(5) \\
$z_{As}$  &0.3538(1)	   &0.6512(2)  \\
Source    &Ref.~\cite{rott}&Ref.~\cite{kamihara_08}\\
\hline
Ba-As, \AA&3.372(1)$\times$8&  -	\\ 
La-As, \AA&-	     &3.380$\times$4	\\ 
Fe-As, \AA&2.388(1)$\times$4&2.412$\times$4	\\ 
Fe-Fe, \AA&2.764(1)$\times$4&2.853$\times$4	\\ 
As-Fe-As  &109.9(1)$^\circ$ &113.6$^\circ$ \\ 
          &109.3(1)$^\circ$ &107.5$^\circ$ \\ 
\hline
\end{tabular}
\end{table}

\begin{figure}
\includegraphics[clip=true,width=0.9\columnwidth,angle=270]{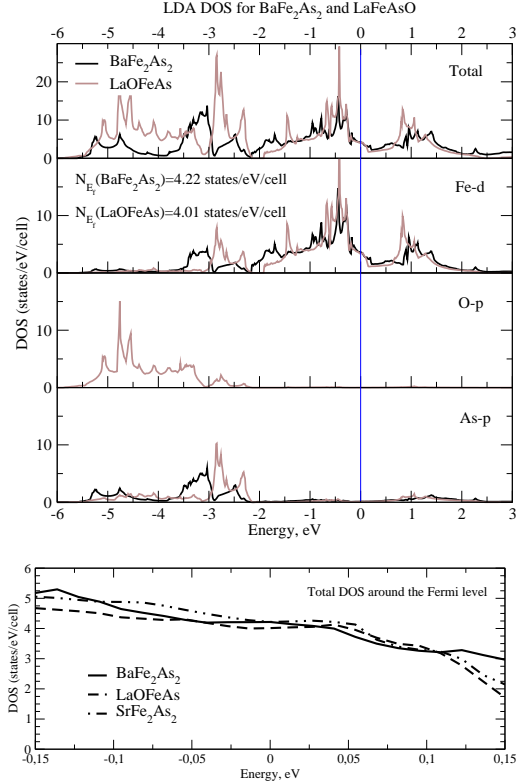}
\includegraphics[clip=true,width=0.8\columnwidth]{total_dos_comp.eps}
\caption{\label{la_ba_dos} 
Fig. 2. Total and partial LDA DOS for BaFe$_2$As$_2$ 
(black lines) and LaOFeAs (light lines) compounds.
Lower panel presents total DOS for
different FeAs systems in the vicinity of the Fermi level.
The Fermi level corresponds to zero. }
\end{figure}

 
The electronic structure of BaFe$_2$As$_2$ and LaOFeAs compounds was calculated 
within the local density approximation (LDA) by using linearized muffin-tin orbitals basis 
(LMTO)~\cite{LMTO}.
For BaFe$_2$As$_2$ we used the structure data for K-doped system and temperature 
T=20 K~\cite{rott}.
The LDA calculated total and partial densities of states for BaFe$_2$As$_2$ and 
LaOFeAs are shown in Fig.~\ref{la_ba_dos}.  
In the lower panel of Fig.~\ref{la_ba_dos} we show magnified behavior of total 
DOS around the Fermi level for three different systems under discussion. 
In all cases DOS is almost flat.
It is well known that DOS of two dimensional (nearly free) electrons is a 
constant defined just by the renormalized electron mass.
Thus, our results support the two-dimensional nature of these compounds.

The values of density of states at Fermi level are very similar 
in both compounds. The 0.3 eV wider Fe-$d$ bandwidth in 
the case of BaFe$_2$As$_2$ in comparison with LaOFeAs arises from the shorter 
Fe-As bonds and hence stronger Fe-$d$-As-$p$ hybridization for this system. 
The partial As-$p$ DOS is splitted into two parts in the case of Ba system. 
The orbital projected Fe-3$d$ DOS for two compounds is 
shown in Fig.~3.
One can see that for both systems three Fe-$d$ orbitals of $t_{2g}$ symmetry -- 
$xz$, $yz$, $x^2-y^2$ mainly 
contribute to the bands crossing the Fermi level.
We call here the $x^2-y^2$ (basically rotated $xy$ orbital) as one of $t_{2g}$
orbitals following the established earlier terminology for ReOFeAs systems.

Energy bands along the high symmetry directions of the 
Brillouin zone are pictured in 
Fig.~\ref{la_ba_bands}. The bands around the Fermi level for both compounds are primarily 
formed by Fe-$d$ states. In LaOFeAs system As-$p$ states are also hybridized
with O-$p$ states and the corresponding bands are separated from the Fe-$d$ ones. 
On the contrary in BaFe$_2$As$_2$ Fe-$d$ and As-$p$ bands are entangled.
The lower two panels of Fig.~\ref{la_ba_bands} compare band dispersions
for both system close to the Fermi level.
Here only $k_x, k_y$ disperion is shown. Taking into account different notations
of high-symmetry points for these two different crystal structures one can find 
these dispersions pretty similar to each other. There are three hole-like bands 
around $\Gamma$-point and two electron bands around $X$-point.  
Thus one can define a minimal model
of ``bare'' electronic bands to treat e.g. superconductivity, similar to
that discussed in Ref. \cite{Gork}. 
Let us mention that along $X-M$ direction in LaOFeAs
there are two degenerate bands.

In Fig. 5 relative on-site energies of hybridized Fe-3$d$ and As-4$p$ states are presented.
A bird's eye view tells us that this picture for 
both BaFe$_2$As$_2$ (left) and LaOFeAs (right) is rather similar. There are 
two groups of states -- antibonding (mostly Fe-3$d$) and bonding (mostly 
As-4$p$) states.  However, there are some fine differences. First of all for 
BaFe$_2$As$_2$ hybridization between Fe-3$d$-$z^2$ and As-4$p_z$ orbitals is 
about 0.24 eV weaker. It leads to a swap of the energy positions of 
Fe-3$d$-$z^2$ and $x^2$ orbitals and similarly for corresponding As-4$p$ 
orbitals.  Secondly, Fe-$d$-$t_{2g}$ orbitals are degenerate
for BaFe$_2$As$_2$ in contrast to LaOFeAs.

Neglecting small difference, the overall picture of the energy spectrum in the
vicinity of the Fermi level is very similar for both compounds
and is determined mainly by Fe-$d$ states of FeAs layers, making the states of A-ions
or rare-earths Re more or less irrelevant for superconductivity. 
Thus, superconductivity of FeAs layered compounds may be studied within the
minimal model, taking into account only essential Fe-$d$ bands close to the   
Fermi level. The variants of such model proposed e.g. in Refs. \cite{Gork,Scal}
for LaOFeAs system may also be used for AFe$_2$As$_2$ with only slight 
modification of model parameters, such as transfer integrals.

The role of electronic correlations in AFe$_2$As$_2$ and ReOFeAs compounds
remains at the moment disputable.
On general grounds it can be expected to be rather important due to large
values of Hubbard and Hund interactions on Fe. However, LDA+DMFT calculations
for LaOFeAs reported in
Refs. \cite{haule,anis} have produced rather contradictory claims.
Obviously, this problem requires further studies. Assuming that correlations
in these compounds are most likely in the intermediate range, we may hope
that standard LDA approach used here is reliable enough.

\begin{figure}
\includegraphics[clip=true,width=0.9\columnwidth,angle=270]{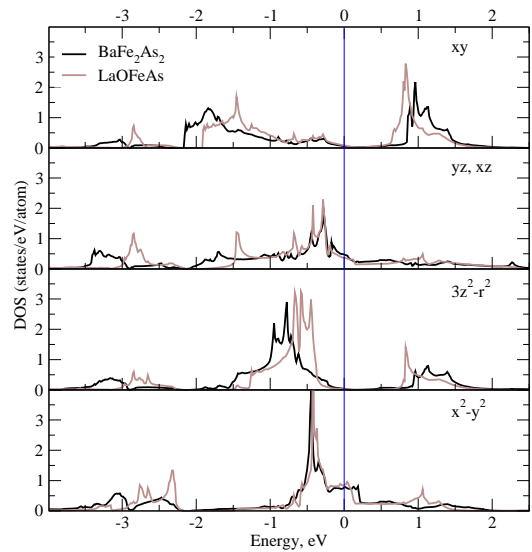}
\caption{\label{la_ba_pdos}Fig. 3. Orbital projected Fe-$d$ DOS for BaFe$_2$As$_2$ 
(black lines) and 
LaOFeAs (light lines) compounds.
The Fermi level corresponds to zero.}
\end{figure}

\begin{figure}
\includegraphics[clip=true,width=0.6\columnwidth,angle=270]{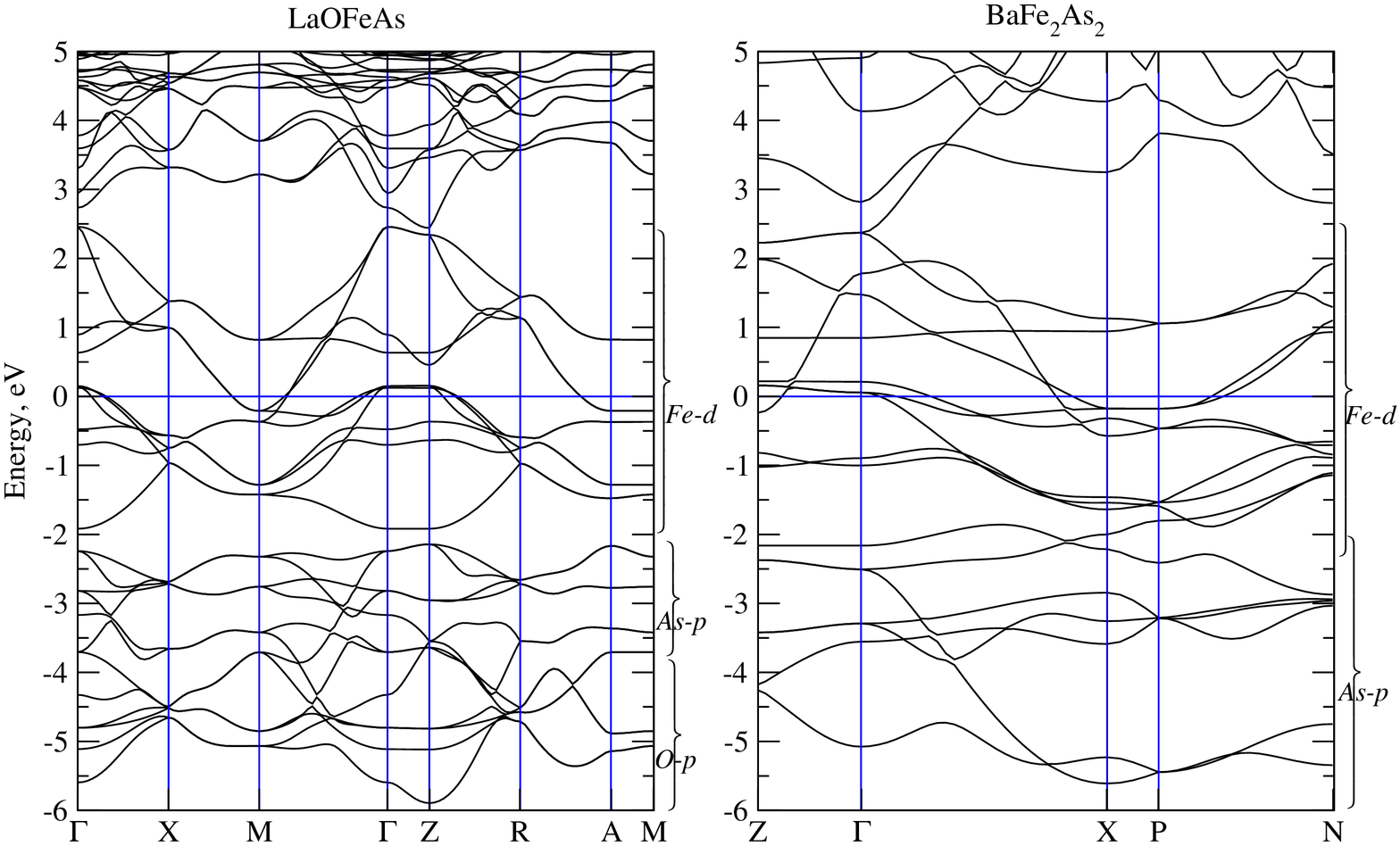}
\includegraphics[clip=true,width=0.8\columnwidth]{bands_comp.eps}

\caption{\label{la_ba_bands} 
Fig. 4. Energy bands for LaOFeAs (left) and BaFe$_2$As$_2$ (right) compounds.
Lower two panels present $k_x,k_y$ disperions for BaFe$_2$As$_2$ and LaOFeAs systems
in the vicinity of the Fermi level.
The Fermi level corresponds to zero.}
\end{figure}

\begin{figure}[!h]
\includegraphics[clip=true,width=0.3\textwidth,angle=270]{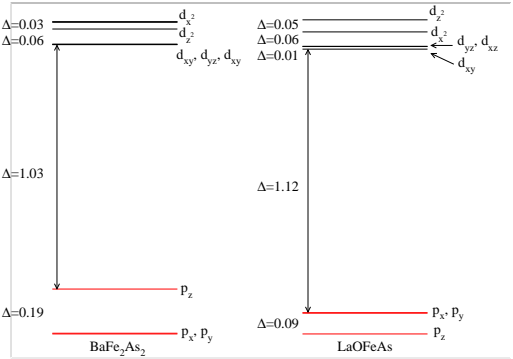}
\label{cf}
\caption{Fig. 5. Relative on-site energies of hybridized Fe-3$d$ and As-4$p$ states obtained 
from LDA dispersions for BaFe$_2$As$_2$ (left) and LaOFeAs 
(right).  $\Delta$ stands for the corresponding energy distances in eV.} 
\end{figure}

\begin{figure}
\includegraphics[clip=true,width=0.8\columnwidth]{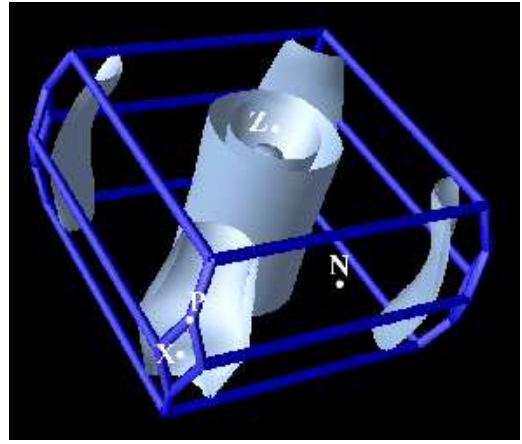}
\caption{\label{ba_fs} Fig. 6. Fermi surface of BaFe$_2$As$_2$
shown in the first Brillouin zone centered at $\Gamma$ point.}
\end{figure}

\begin{figure}
\includegraphics[clip=true,width=0.8\columnwidth]{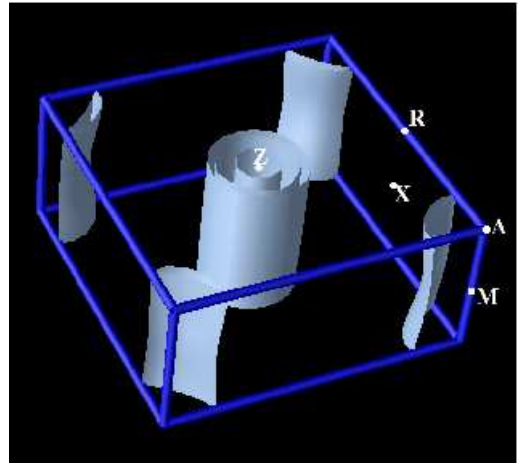}
\caption{\label{la_fs} Fig. 7.
The same as Fig. 6 but for LaOFeAs.}
\end{figure}

Fermi surfaces obtained from LDA calculations for BaFe$_2$As$_2$ and LaOFeAs
are shown in Figs. 6 and 7, correspondingly. There are 
five sheets of Fermi surface for both compounds. Qualitatively, Fermi 
surfaces are similar to that reported for LaOFeAs in Ref. \cite{singh} 
(see also \cite{mazin}).  There are three hole cylinders in the middle of the 
Brillouin zones and two electron sheets at the corners of Brillouin zone. 
Smallest of hole cylinders is usually neglected in the analysis of 
superconducting pairings \cite{Gork, Scalap} and analysis is restricted to 
minimal two \cite{Scalap} or four bands \cite{Gork} models, reproducing two 
hole and two electron cylinders.

$P$4/$nmm$ (tetragonal)$\rightarrow$ $Cmma$ (orthorhombic) phase transition
taking place in undoped compounds is usually attributed to SDW formation due
to nesting properties of electron and hole Fermi surfaces \cite{mazin,Scal} or
due to excitonic instability in triplet channel \cite{Gork}. The difficulties
of calculating magnetic state of LaOFeAs related with apparently itinerant nature of
magnetism were recently discussed in Ref. \cite{maz}. 

In conclusion, we have presented the results of LDA calculations of new
prototype high-temperature superconductor AFe$_2$As$_2$ (A=Ba, Sr) and compared 
it with previously discussed ReOFeAs series, demonstrating essential similarity of
electronic states close to the Fermi level and most important for 
superconductivity. These states are formed mainly by Fe orbitals in the 
two-dimensional FeAs layer, which is the basic structural motif where superconducting state
is formed. Thus, rather simple AFe$_2$As$_2$ system may be considered 
generic for the studies of high-temperature superconductivity in whole class of 
FeAs-layered compounds.

This work is supported by RFBR grants 08-02-00021, 08-02-00712, RAS programs 
``Quantum macrophysics'' and ``Strongly correlated electrons in 
semiconductors, metals, superconductors and magnetic materials'',
Grants of President of Russia MK-2242.2007.2(IN), MK-3227.2008.2(ZP)
and scientific school grant SS-1929.2008.2,  interdisciplinary 
UB-SB RAS project, Dynasty Foundation (ZP) and Russian Science Support 
Foundation(IN). The authors are grateful to L.P. Gorkov for useful discussions.


\begin{thebibliography}{99}

\bibitem{kamihara_08} Y. Kamihara, T. Watanabe, M. Hirano, H. Hosono, J. Am. Chem. Soc. {\bf 130}, 3296-3297 (2008).

\bibitem{chen}G.F. Chen, Z. Li, G. Zhou, D. Wu, J. Dong, W.Z. Hu, P. Zheng, 
Z.J. Chen, J.L. Luo, N.L. Wang, arXiv: 0803.0128.

\bibitem{zhu}X. Zhu, H. Yang, L. Fang, G. Mu, H.-H. Wen, arXiv: 0803.1288v1.

\bibitem{mand}A.S. Sefat, M.A. McGuire, B.C. Sales, R. Jin, J.Y. Hove, 
D. Mandrus, arXiv: 0803.2528.

\bibitem{chen_3790} G. F. Chen, Z. Li, D. Wu, G. Li, W. Z. Hu, J. Dong, P. Zheng, J. L. Luo, N. L. Wang, arXiv:0803.3790.

\bibitem{chen_3603} X. H. Chen, T. Wu, G. Wu, R. H. Liu, H. Chen, D. F. Fang, arXiv:0803.3603.

\bibitem{ren_4234} Z.-A. Ren, J. Yang, W. Lu, W. Yi, X.-L Shen, G.-C. Che, 
L.-L. Sun, F. Zhou, Z.-X. Zhao, arXiv:0803.4234.

\bibitem{ren_4283} Z.-A. Ren, J. Yang, W. Lu, W. Yi, G.-C. Che, X.-Li. Dong, 
L.-L. Sun, Z.-X. Zhao, arXiv:0803.4283.

\bibitem{rott}M. Rotter, M. Tegel, D. Johrendt, arXiv: 0805.4630.

\bibitem{ChenLi}G.F. Chen, Z. Li, G. Li, W.Z. Hu, J. Dong, X.D. Zhang,
N.L. Wang, J.L. Luo, arXiv: 0806.1209.

\bibitem{Chu}K. Sasmal, B. Lv, B. Lorenz, A. Guloy, F. Chen, Y. Xue, C.W. Chu,
arXiv: 0806.1301.

\bibitem{Bud}N. Ni, S.L. Bud'ko, A. Kreyssig, S. Nandi, G.E. Rustan, A.I. Goldman,
S. Gupta, J.D. Corbett, A. Kracher, P.C. Canfield, arXiv: 0806.1874.

\bibitem{singh} D.J. Singh, M.H. Du. Phys. Rev. Lett. {\bf 100}, 237003 (2008), 
arXiv:0803.0429.

\bibitem{dolg}L. Boeri, O.V. Dolgov, A.A. Golubov, arXiv: 0803.2703v1.

\bibitem{mazin}I.I. Mazin, D.J. Singh, M.D. Johannes, M.H. Du,
arXiv: 0803.2740v1.

\bibitem{lebegue}S. Leb\`{e}gue, Phys. Rev. B 75, 035110 (2007). 

\bibitem{Nekr}I.A. Nekrasov, Z.V. Pchelkina, M.V. Sadovskii. Pis'ma Zh. Eksp.
Teor. Fiz. {bf 87}, 647 (2008) [JETP Letters {\bf 87} (2008)],
arXiv: 0804.1239.

\bibitem{Shein}I.R. Shein, A.L. Ivanovskii, arXiv: 0806.0750.

\bibitem{Krell}C. Krellner, N. Caroca-Canales, A. Jesche, H. Rosner, A. Ormeci,
C. Geibel, arXiv: 0806.1043.

\bibitem{rotter_4021} M. Rotter, M. Tegel, D. Johrendt, arXiv: 0805.4021.

\bibitem{Nomura_3569} T. Nomura, S. W. Kim, Y. Kamihara, M. Hirano, P. V. Sushko, K. Kato, M. Takata, A. L. Shluger, 
H. Hosono, arXiv: 0804.3569.


\bibitem{LMTO}O.K. Andersen. Phys. Rev. B {\bf 12} 3060 (1975);
O. Gunnarsson, O. Jepsen,  O.K. Andersen. Phys. Rev. B {\bf 27} 7144 (1983);
O.K. Andersen, O. Jepsen.  Phys. Rev. Lett. {\bf 53} 2571  (1984).

\bibitem{Gork}V. Barzykin, L.P. Gorkov, arXiv: 0806.1993, JETP Letters (to be
published).


\bibitem{Scal}S. Raghu, Xiao-Liang Qi, Chao-Xing Liu, D.J. Scalapino, 
Shou-Cheng Zhang, arXiv: 0804.1113.

\bibitem{haule}K. Haule, J.H. Singh, G. Kotliar. Phys. Rev. Lett. {\bf 100},
226402 (2008), arXiv: 0803.1279.

\bibitem{anis}A.O. Shorikov, M.A. Korotin, S.V. Streltsov, D.M. Korotin,
V.I. Anisimov, S.L. Skornyakov, arXiv: 0804.3283.

\bibitem{Scalap}Xiao-Liang Qi, S. Raghu, Chao-Xing Liu, D.J. Scalapino, 
Shou-Cheng Zhang, arXiv: 0804.4332.

\bibitem{maz}I.I. Mazin, M.D. Johannes, L. Boeri, K. Koepernik, D.J. Singh,
arXiv: 0806.1869.


\end{thebibliography}
\end{document}